\documentstyle[12pt]{article}
\begin{document}

\title{\textbf{$f(T)$ modified teleparallel gravity models as an alternative
for holographic and new agegraphic dark energy models}}

\author{K. Karami$^{1,2}$\thanks{E-mail: KKarami@uok.ac.ir} ,
A. Abdolmaleki${^1}$\thanks{E-mail:
AAbdolmaleki@uok.ac.ir}\\$^{1}$\small{Department of Physics,
University of Kurdistan, Pasdaran St., Sanandaj,
Iran}\\$^{2}$\small{Research Institute for Astronomy
$\&$ Astrophysics of Maragha (RIAAM), Maragha, Iran}\\
}

\maketitle

\begin{abstract}
In the present work, we reconstruct different $f(T)$-gravity models
corresponding to the original and entropy-corrected version of the
holographic and new agegraphic dark energy models. We also obtain
the equation of state parameters of the corresponding $f(T)$-gravity
models. We conclude that the holographic and new agegraphic
$f(T)$-gravity models behave like phantom or quintessence model.
Whereas in the entropy-corrected models, the equation of state
parameter can justify the transition from the quintessence state to
the phantom regime as indicated by the recent observations.
\end{abstract}
\clearpage
\section{Introduction}
Recent observational data coming from type Ia supernovae (SNeIa)
surveys, large scale structure (LSS), and the cosmic microwave
background (CMB) anisotropy spectrum points toward the picture of a
spatially flat universe undergoing an accelerated expansion driven
by a dominant negative pressure fluid, typically referred to as dark
energy (DE) \cite{Riess}. It is shown that DE takes up about
two-thirds of the total energy density from cosmic observations.
Although the nature and cosmological origin of DE is still enigmatic
at the present, a great variety of models have been proposed to
describe the DE (for review see \cite{Padmanabhan,Copeland}). Two
promising candidates are the holographic DE (HDE) \cite{Li} and the
agegraphic DE (ADE) \cite{Cai} models which are originated from some
considerations of the features of the quantum theory of gravity.

The HDE model is motivated from the holographic principle
\cite{Suss1}. Following Guberina et al. \cite{Guberina}, the HDE
density can be derived from the entropy bound. In the thermodynamics
of the black hole \cite{Bekenstein}, there is a maximum entropy in a
box of size $L$, namely, the Bekenstein-Hawking entropy bound
$S_{\rm BH}\sim M_P^2 L^2$, which scales as the area of the box $A
\sim L^2$, rather than the volume $V \sim L^3$. Here $M_P$ is the
reduced Planck Mass $M_P^{-2}=8\pi G$. Also for a macroscopic system
in which self-gravitation effects can be disregarded, the Bekenstein
entropy bound $S_{\rm B}$ is given by the product of the energy
$E\sim\rho_{\Lambda}L^3$ and the length scale (IR cut-off) $L$ of
the system. Here $\rho_{\Lambda}$ is the quantum zero point energy
density caused by the UV cut-off $\Lambda$. Requiring $S_{\rm B}\leq
S_{\rm BH}$, namely $EL\leq M_P^2 L^2$, one has $\rho_{\Lambda}\leq
M_P^2L^{-2}$. If the largest cut-off $L$ is taken for saturating
this inequality, we get the energy density of the HDE as
\begin{equation}
\rho_{\Lambda}=3c^2M_P^2L^{-2},\label{rhoHDE}
\end{equation}
where $c$ is a numerical constant. Recent observational data, which
have been used to constrain the HDE model, show that for the
non-flat universe $c=0.815_{-0.139}^{+0.179}$ \cite{Li5}, and for
the flat case $c=0.818_{-0.097}^{+0.113}$ \cite {Li6}. Li \cite{Li}
showed that the cosmic coincidence problem can be resolved by
inflation in the HDE model, providing the minimal number of
e-foldings. The HDE models have been studied widely in the
literature \cite{Enqvist,Elizalde2,Guberina1,Guberina2,Karami1}.
Indeed, the definition and derivation of the HDE density depends on
the entropy-area relationship $S_{\rm BH} = A/(4G)$, where $A\sim
L^2$ is the area of horizon. However, this definition can be
modified by the inclusion of quantum effects, motivated from the
loop quantum gravity (LQG). These quantum corrections provided to
the entropy-area relationship lead to the curvature correction in
the Einstein-Hilbert action and vice versa \cite{Zhu}. The corrected
entropy takes the form \cite{modak}
\begin{equation}
S_{\rm
BH}=\frac{A}{4G}+\tilde{\alpha}\ln{\frac{A}{4G}}+\tilde{\beta},\label{MEAR}
\end{equation}
where $\tilde{\alpha}$ and $\tilde{\beta}$ are dimensionless
constants of order unity. Determination of the exact values of these
constants is still an open issue in quantum gravity. These
corrections arise in the black hole entropy in the LQG due to
thermal equilibrium fluctuations and quantum fluctuations
\cite{Rovelli}. Taking the corrected entropy-area relation
(\ref{MEAR}) into account, and following the derivation of the HDE
(especially the one shown in \cite{Guberina}), the energy density of
the HDE will be modified. On this basis, Wei \cite{HW} proposed
 the energy density of the so-called ``entropy-corrected HDE''
 (ECHDE) in the form
\begin{equation}
\rho_{\Lambda}=3c^2M_{P}^{2}L^{-2}+\alpha L^{-4}\ln
(M_{P}^{2}L^{2})+\beta L^{-4},\label{rhoECHDE}
\end{equation}
where $\alpha$ and $\beta$ are dimensionless constants of order
unity. In the special case $\alpha=\beta=0$, the above equation
yields the well-known HDE density (\ref{rhoHDE}). Since the last two
terms in Eq. (\ref{rhoECHDE}) can be comparable to the first term
only when $L$ is very small, the corrections make sense only at the
early stage of the universe. When the universe becomes large, the
ECHDE reduces to the ordinary HDE model. The ECHDE models have
arisen a lot of enthusiasm recently and have examined in ample
detail by \cite{Khodam}.

The ADE model is originated from the uncertainty relation of quantum
mechanics together with the gravitational effect in general
relativity (GR). The ADE model assumes that the observed DE comes
from the spacetime and matter field fluctuations in the universe.
Following the line of quantum fluctuations of spacetime, Karolyhazy
et al. \cite{Kar1} discussed that the distance $t$ in Minkowski
spacetime cannot be known to a better accuracy than
$\delta{t}\propto t_{P}^{2/3}t^{1/3}$, where $t_P$ is the reduced
Planck time. Based on the Karolyhazy relation, Maziashvili
\cite{Maz} discussed that the energy density of the metric
fluctuations of the Minkowski spacetime is given by
$\rho_{\Lambda}\sim 1/(t_Pt)^2\sim M_P^2t^{-2}$. Based on the
Karolyhazy relation \cite{Kar1} and Maziashvili arguments
\cite{Maz}, Cai proposed the original ADE model to explain the
accelerated expansion of the universe \cite{Cai}. The original ADE
has the energy density $\rho_{\Lambda}=3{n}^2M_P^2 T^{-2}$, where
$T$ is the age of the universe \cite{Cai}. Also the numerical factor
3$n^2$ is introduced to parameterize some uncertainties, such as the
species of quantum fields in the universe, the effect of curved
spacetime (since the energy density is derived for Minkowski
spacetime), and so on. However, the original ADE model had some
difficulties. For example it suffers from the difficulty to describe
the matter-dominated epoch. Therefore, a new model of the ADE was
proposed by Wei and Cai \cite{Wei1}, while the time scale is chosen
to be the conformal time instead of the age of the universe. The
energy density of the new ADE (NADE) is given by \cite{Wei1}
\begin{equation}
\rho_{\Lambda}=3{n}^2M_P^2 \eta^{-2},\label{rhoNADE}
\end{equation}
where $\eta$ is the conformal time of the FRW universe. The joint
analysis of the astronomical data for the NADE model in flat
universe gives the best-fit value (with 1$\sigma$ uncertainty)
$n=2.716_{-0.109}^{+0.111}$ \cite{Wei3}. It was found that the
coincidence problem could be solved naturally in the NADE model
\cite{Wei3}. The ADE models have been examined and studied in ample
detail by \cite{Wei2,Kim,Sheykhi}. More recently, very similar to
the ECHDE model, the energy density of the entropy-corrected NADE
(ECNADE) was proposed by Wei \cite{HW} as
\begin{equation}
\rho_{\Lambda} = 3n^2{M_P^2}\eta^{-2} +
\alpha\eta^{-4}\ln{({M_P^2}{\eta}^2)} +
\beta\eta^{-4}.\label{rhoECNADE}
\end{equation}
In the special case $\alpha=\beta=0$, Eq. (\ref{rhoECNADE}) yields
the NADE density (\ref{rhoNADE}). The motivation for taking the
energy density of the modified NADE in the form (\ref{rhoECNADE})
comes from the fact that both the NADE and HDE models have the same
origin. Indeed, it was argued that the NADE models are the HDE model
with different IR length scales \cite{Myung}. The ECNADE has been
investigated in ample detail in \cite{Karami2}.

One of among other interesting alternative proposals for
 DE is modified gravity. It can explain naturally the
unification of earlier and later cosmological epochs (for review see
\cite{Capozziello}). Moreover, modified gravity may serve as dark
matter \cite{Sobouti}. There are some classes of modified gravities
containing $f(R)$, $f(\mathcal{G})$ and $f(R,\mathcal{G})$ which are
considered as gravitational alternatives for DE
\cite{NojiriOdin,husawiski,noj abdalla,Noj2}. Here the Lagrangian
density of modified gravity theories $f$ is an arbitrary function of
$R$, $\mathcal{G}$ or both $R$ and $\mathcal{G}$. The field
equations of these modified gravity theories are the 4th order that
making it difficult obtain both exact and numerical solutions.
Recently, a new modified gravity model was proposed by Bengochea and
Ferraro \cite{bengochea} to describe the present accelerating
expansion of the universe without resorting to
 DE. Instead of using the curvature defined via
the Levi-Civita connection in GR, the Weitzenb\"{o}ck connection is
used in teleparallel gravity \cite{Einstein}. As a result, the
spacetime has no curvature but contains torsion. Similar to GR where
the action is a curvature scalar $R$, the action of teleparallel
gravity is a torsion scalar $T$. Following this line and in analogy
to the $f(R)$ theory, Bengochea and Ferraro \cite{bengochea}
suggested a new model, named $f(T)$ theory, by generalizing the
action of teleparallel gravity as a function of the torsion scalar
$T$, and found that it can explain the observed acceleration of the
universe. Indeed, there are some terms in the modified Friedmann
equation in $f(T)$-gravity that can be identified as the effective
DE to produce the accelerated expansion of the late-time universe
\cite{Bamba,yerzhanov}. Models based on modified teleparallel
gravity may also provide an alternative to inflation \cite{Ferraro}.
Another advantage of $f(T)$ theory is that its field equations are
the second order which are remarkably simpler than the fourth order
equations of $f(R)$ theory \cite{WufT}. Recently, $f(T)$-gravity has
been extensively studied in the literature
\cite{Bamba,yerzhanov,WufT,linder,Dent}.

Viewing the $f(T)$ modified gravity model as an effective
description of the underlying theory of DE, and considering the
original and entropy-corrected version of the HDE and NADE scenarios
as pointing in the direction of the underlying theory of
 DE, it is interesting to study how the $f(T)$-gravity
can describe the HDE, ECHDE, NADE and ECNADE densities as effective
theories of DE models. This motivated us to establish different
models of $f(T)$-gravity according to the original and
entropy-corrected version of the HDE and NADE scenarios. This paper
is organized as follows. In section 2, we review the theory of
$f(T)$-gravity in the metric formalism. In sections 3, 4, 5 and 6,
we reconstruct different $f(T)$-gravity models corresponding to the
HDE, ECHDE, NADE and ECNADE models, respectively. Section 7 is
devoted to our conclusions.

\section{$f(T)$ modified teleparallel gravity}

In the framework of $f(T)$ theory, the action of modified
teleparallel gravity is given by \cite{bengochea}
\begin{equation}
I =\frac{1}{2k^2}\int {\rm d}^4
x~e~\Big[f(T)+L_m\Big],\label{action}
\end{equation}
where $k^2=M_P^{-2}=8\pi G$ and $e={\rm det}(e^i_{\mu})=\sqrt{-g}$.
Also $T$ and $L_m$ are the torsion scalar and the Lagrangian density
of the matter inside the universe, respectively. Note that
$e^i_{\mu}$ is the vierbein field which is used as a dynamical
object in teleparallel gravity and has the following orthonormal
property \cite{bengochea}
\begin{equation}
{\bf e}_i\cdot{\bf e}_j=\eta_{ij},
\end{equation}
where $\eta_{ij}={\rm diag}(1,-1,-1,-1)$. Each vector ${\bf e}_i$
can be described by its components $e_i^{\mu}$, where $i=0,1,2,3$
refers to the tangent space of the manifold and $\mu=0,1,2,3$ labels
coordinates on the manifold. The metric tensor is obtained from the
dual vierbein as
\begin{equation}
g_{\mu\nu}(x)=\eta_{ij}e^i_{\mu}(x)e^j_{\nu}(x).\label{gmetric}
\end{equation}
The torsion scalar $T$ is defined as \cite{bengochea}
\begin{equation}
T =S^{~\mu\nu}_{\rho}T^{\rho}_{~\mu\nu},\label{T1}
\end{equation}
where the non-null torsion tensor $T^{\rho}_{~\mu\nu}$ is
\begin{equation}
T^{\rho}_{~\mu\nu}=e^{\rho}_{i}(\partial_{\mu}e^{i}_{\nu}-\partial_{\nu}e^{i}_{\mu}),
\end{equation}
and
\begin{equation}
S_{\rho}^{~\mu\nu}=\frac{1}{2}(K^{\mu\nu}_{~~~\rho}+\delta^{\mu}_{\rho}T^{\alpha\nu}_{~~~\alpha}-\delta^{\nu}_{\rho}
T^{\alpha\mu}_{~~~\alpha}).
\end{equation}
Also $K^{{\mu\nu}}_{~~~\rho}$ is the contorsion tensor defined as
\begin{equation}
K^{{\mu\nu}}_{~~~\rho}=-\frac{1}{2}(T^{\mu\nu}_{~~~\rho}-T^{\nu\mu}_{~~~\rho}-T_{\rho}^{~\mu\nu}).
\end{equation}
Taking the variation of the action (\ref{action}) with respect to
the vierbein, one can obtain the field equations in $f(T)$ modified
teleparallel gravity as \cite{bengochea}
\begin{equation}
S_i^{~\mu\nu}\partial_{\mu}
(T)f_{TT}(T)+\Big[e^{-1}\partial_{\mu}(eS_i^{~\mu\nu})-e_i^{\lambda}T^{\rho}_{~\mu\lambda}S_{\rho}^{~\nu\mu}\Big]f_T(T)
+\frac{1}{4}e_i^{\nu}f(T)=\frac{k^2}{2}e_i^{~\rho}T_{\rho}^{~\nu},\label{fTeqs}
\end{equation}
where subscript $T$ denotes a derivative with respect to $T$,
$S_i^{~\mu\nu}=e_i^{~\rho}S_{\rho}^{~\mu\nu}$ and $T_{\mu\nu}$ is
the matter energy-momentum tensor. The set of field equations
(\ref{fTeqs}) are the 2nd order which makes them simpler than the
corresponding field equations in the other modified gravity theories
like $f(R)$, $f(\mathcal{G})$ and $f(R,\mathcal{G})$
\cite{yerzhanov}.

Now if we consider the spatially-flat FRW metric for the universe as
\begin{equation}
g_{\mu\nu}={\rm diag}\Big(-1,a^2(t),a^2(t),a^2(t)\Big),
\end{equation}
where $a$ is the scale factor, then from Eq. (\ref{gmetric}) one can
obtain
\begin{equation}
e^{i}_{\mu}={\rm diag}\Big(1,a(t),a(t),a(t)\Big).\label{e}
\end{equation}
Substituting the vierbein (\ref{e}) into (\ref{T1}) yields
\cite{bengochea}
\begin{equation}
T=-6H^2,\label{T}
\end{equation}
where $H=\dot{a}/a$ is the Hubble parameter.

Taking $T_\nu^\mu={\rm diag}(-\rho,p,p,p)$ for the matter
energy-momentum tensor in the prefect fluid form and using the
vierbein (\ref{e}), the set of field equations (\ref{fTeqs}) for
$i=0=\nu$ reduces to \cite{bengochea}
\begin{equation}
12H^2f_T(T)+f(T)=2k^2\rho,\label{fT1}
\end{equation}
and for $i=1=\nu$ yields
\begin{equation}
48H^2\dot{H}f_{TT}(T)-(12H^2+4\dot{H})f_T(T)-f(T)=2k^2p.\label{fT2}
\end{equation}
Here $\rho$ and $p$ are the total energy density and pressure of the
matter inside the universe, respectively, and satisfy the
conservation equation
\begin{equation}
\dot{\rho}+3H(\rho+p)=0.\label{conteq}
\end{equation}
Note that Eqs. (\ref{fT1}) and (\ref{fT2}) are the modified
Friedmann equations in the framework of $f(T)$-gravity in the
spatially-flat FRW universe. One can rewrite Eqs. (\ref{fT1}) and
(\ref{fT2}) as \cite{yerzhanov}
\begin{equation}
\frac{3}{k^2}H^2=\rho+\rho_T,\label{fT11}
\end{equation}
\begin{equation}
\frac{1}{k^2}(2\dot{H}+3H^2)=-(p+p_T),\label{fT22}
\end{equation}
where
\begin{equation}
\rho_T=\frac{1}{2k^2}(2Tf_T-f-T),\label{roT}
\end{equation}
\begin{equation}
p_T=-\frac{1}{2k^2}[-8\dot{H}Tf_{TT}+(2T-4\dot{H})f_T-f+4\dot{H}-T],\label{pT}
\end{equation}
are the torsion contribution to the energy density and pressure
which satisfy the energy conservation law
\begin{equation}
\dot{\rho}_T+3H(\rho_T+p_T)=0.\label{EfT}
\end{equation}
In the case of $f(T)=T$, from Eqs. (\ref{roT}) and (\ref{pT}) we
have $\rho_T=0$ and $p_T=0$. Therefore, Eqs. (\ref{fT11}) and
(\ref{fT22}) are transformed to the usual Friedmann equations in GR.

The equation of state (EoS) parameter due to the torsion
contribution is defined as
\begin{equation}
\omega_T=\frac{p_T}{\rho_T}=-1+\frac{4\dot{H}(2Tf_{TT}+f_T-1)}{2Tf_T-f-T}.\label{omegaT}
\end{equation}
Note that for the de Sitter universe, i.e. $\dot{H}=0$, we have
$\omega_T=-1$ which behaves like the cosmological constant. Also for
a $f(T)$-dominated universe, Eq. (\ref{fT11}) yields
\begin{equation}
\frac{3}{k^2}H^2=\rho_T.
\end{equation}
Taking time derivative of the above relation and using the
continuity equation (\ref{EfT}), one can get the EoS parameter as
\begin{equation}
\omega_{T}=-1-\frac{2\dot{H}}{3H^2},
\end{equation}
which shows that for the phantom-dominated, $\omega_{T}<-1$, and
quintessence-dominated, $\omega_{T}>-1$, universe, we need to have
$\dot{H}>0$ and $\dot{H}<0$, respectively.

For a given $a=a(t)$, with the help of Eqs. (\ref{roT}) and
(\ref{pT}) one can reconstruct the $f(T)$-gravity according to any
DE model given by the EoS $p_T=p_T(\rho_T)$ or $\rho_T=\rho_T(a)$.
There are two classes of scale factors which usually people consider
for describing the accelerating universe in
 $f(R)$, $f(\mathcal{G})$ and $f(R,\mathcal{G})$
modified gravities \cite{Nojiri}.

The first class of scale factors is given by \cite{Nojiri,Sadjadi}
\begin{equation}
a(t)=a_0(t_s-t)^{-h},~~~t\leq t_s,~~~h>0.\label{a}
\end{equation}
Using Eqs. (\ref{T}) and (\ref{a}) one can obtain
\begin{equation}
H=\frac{h}{t_s-t},~~~T=-\frac{6h^2}{(t_s-t)^2},~~~
\dot{H}=-\frac{T}{6h},\label{respect to r}
\end{equation}
in which the last relation $\dot{H}=-T/(6h)=H^2/h>0$ shows that the
model (\ref{a}) is corresponding to a phantom-dominated universe.
This is why in the literature the model (\ref{a}) is usually
so-called the phantom scale factor.

For the second class of scale factors defined as \cite{Nojiri}
\begin{equation}
a(t)=a_0t^h,~~~h>0,\label{aQ}
\end{equation}
one can obtain
\begin{equation}
H=\frac{h}{t},~~~T=-\frac{6h^2}{t^2},~~~
\dot{H}=\frac{T}{6h},\label{respect to rQ}
\end{equation}
in which the last relation $\dot{H}=T/(6h)=-H^2/h<0$ reveals that
the model (\ref{aQ}) describes a quintessence-dominated universe.
Due to this fact, the model (\ref{aQ}) is so-called the quintessence
scale factor in the literature.

In the next sections, using the two classes of scale factors
(\ref{a}) and (\ref{aQ}), we reconstruct different $f(T)$-gravities
according to the HDE, ECHDE, NADE and ECNADE models.
\section{Holographic $f(T)$-gravity model}

Here we reconstruct the $f(T)$-gravity from the HDE model. Note that
in Eq. (\ref{rhoHDE}), taking $L$ as the size of the current
universe, for instance, the Hubble scale, the resulting energy
density is comparable to the present day DE. However, as Hsu found
in \cite{Enqvist}, in this case, the evolution of the DE is the same
as that of dark matter (dust matter), and therefore it cannot drive
the universe to accelerated expansion. The same appears if one
chooses the particle horizon of the universe as the length scale $L$
\cite{Li}. To obtain an accelerating universe, Li \cite{Li} proposed
that for a flat universe, $L$ should be the future event horizon
$R_{h}$. Following Li \cite{Li} the HDE density with the IR cut-off
$L=R_h$ is given by
\begin{equation}
\rho_{\Lambda}=\frac{3c^2}{k^2R_h^2},\label{ro H}
\end{equation}
where the future event horizon $R_h$ is defined as
\begin{equation}
R_h=a\int_t^{\infty}\frac{{\rm d}t}{a}=a\int_a^{\infty}\frac{{\rm
d}a}{Ha^2}.\label{L0}
\end{equation}
For the first class of scale factors (\ref{a}) and using Eq.
(\ref{respect to r}), the future event horizon $R_h$ yields
\begin{equation}
R_h=a\int_t^{t_s}\frac{{\rm
d}t}{a}=\frac{t_s-t}{h+1}=\frac{h}{h+1}{\left(\frac{-6}{T}\right)}^{1/2}.\label{Rh}
\end{equation}
Inserting Eq. (\ref{Rh}) into (\ref{ro H}) one can get
\begin{equation}
\rho_\Lambda=-\frac{\gamma}{2k^2}T,\label{ro H R}
\end{equation}
where
\begin{equation}
\gamma=c^2{\left(\frac{h+1}{h}\right)}^2.\label{alpha}
\end{equation}
Equating (\ref{roT}) with (\ref{ro H R}), i.e.
$\rho_T=\rho_{\Lambda}$, we obtain the following differential
equation
\begin{equation}
2Tf_T-f+(\gamma-1)T=0.\label{dif eq1}
\end{equation}
Solving Eq. (\ref{dif eq1}) yields the holographic $f(T)$-gravity
model as
\begin{equation}
f(T)=\epsilon T^{1/2}+(1-\gamma)T,\label{fHDE}
\end{equation}
where $\epsilon$ is an integration constant.

Substituting Eq. (\ref{fHDE}) into (\ref{omegaT}) one can
 obtain the EoS parameter of the torsion contribution as
\begin{equation}
\omega_{T}=-1-\frac{2}{3h},~~~h>0,\label{wHDE}
\end{equation}
which is always smaller than $-1$ and corresponds to a phantom
accelerating universe. Recent observational data indicates that the
EoS parameter $\omega_{T}$ at the present lies in a narrow strip
around $\omega_{T} = -1$ and is quite consistent with being below
this value \cite{Copeland}.

For the second class of scale factors (\ref{aQ}) and using Eq.
(\ref{respect to rQ}), the future event horizon $R_h$ reduces to
\begin{equation}
R_h=a\int_t^{\infty}\frac{{\rm
d}t}{a}=\frac{t}{h-1}=\frac{h}{h-1}{\left(\frac{-6}{T}\right)}^{1/2},~~~h>1,\label{RhQ}
\end{equation}
where the condition $h>1$ is obtained due to having a finite future
event horizon. If we repeat the above calculations then we can
obtain the both $f(T)$ and $\omega_T$ corresponding to the HDE for
the second class of scale factors (\ref{aQ}). The result for $f(T)$
is the same as (\ref{fHDE}) where
\begin{equation}
\gamma=c^2{\left(\frac{h-1}{h}\right)}^2.\label{alpha}
\end{equation}
Also the EoS parameter is obtained as
\begin{equation}
\omega_{T}=-1+\frac{2}{3h},~~~h>1,\label{wHDEQ}
\end{equation}
which describes an accelerating universe with the quintessence EoS
parameter, i.e. $\omega_T>-1$. It should be mentioned that for
$h>1$, the EoS parameter (\ref{wHDEQ}) also takes place in the range
of $-1<\omega_T<-1/3$.
\section{Entropy-corrected holographic $f(T)$-gravity model}

The ECHDE density (\ref{rhoECHDE}) with the IR cut-off $L=R_h$
yields
\begin{equation}
\rho_\Lambda=\frac{3c^2}{k^2R_h^2}+\frac{\alpha}{R_h^4}\ln\left(\frac{R_h^2}{k^2}\right)+\frac{\beta}{R_h^4}.\label{ECHDE}
\end{equation}
For the first class of scale factors (\ref{a}), substituting Eq.
(\ref{Rh}) into (\ref{ECHDE}) one can get
\begin{equation}
\rho_\Lambda=-\frac{\gamma}{2k^2}T+
\frac{1}{2k^2}\left[\sigma+\delta
\ln\left(-\frac{\lambda}{T}\right)\right]T^2,\label{ro ECH}
\end{equation}
where
\begin{eqnarray}
\gamma=c^2{\left(\frac{h+1}{h}\right)}^2,
~~\delta=\frac{k^2\alpha}{18}{\left(\frac{h+1}{h}\right)}^4,
\nonumber\\
~~\lambda=\frac{6}{k^2}{\left(\frac{h}{h+1}\right)}^2,
~~\sigma=\frac{k^2\beta}{18}{\left(\frac{h+1}{h}\right)}^4.
\end{eqnarray}
Equating (\ref{roT}) with (\ref{ro ECH}) one can get
\begin{equation}
2Tf_T-f+(\gamma-1)T-\left[\sigma+\delta
\ln\left(-\frac{\lambda}{T}\right)\right]T^2=0.\label{dif eq3}
\end{equation}
Solving the differential equation (\ref{dif eq3}) yields the
entropy-corrected holographic $f(T)$-gravity model as
\begin{equation}
f(T)=\epsilon T^{1/2}+(1-\gamma)T+\frac{1}{3}\left\{\sigma+\delta
\left[\frac{2}{3}+\ln\left(-\frac{\lambda}{T}\right)\right]\right\}T^2,\label{fECHDE}
\end{equation}
where $\epsilon$ is an integration constant.

Substituting Eq. (\ref{fECHDE}) into (\ref{omegaT}) one can get
\begin{equation}
\omega_T=-1-\frac{2}{3h}\left[1+
\left({\frac{\delta-[\sigma+\delta\ln{(-\frac{\lambda}{T})}]}
{\gamma-[\sigma+\delta\ln{(-\frac{\lambda}{T})}]T}}\right)T\right],~~~h>0.\label{wECHDE}
\end{equation}
If we set $\delta=0=\alpha$ and $\sigma=0=\beta$ then Eqs.
(\ref{fECHDE}) and (\ref{wECHDE}) reduce to (\ref{fHDE}) and
(\ref{wHDE}), respectively.

\begin{figure}
\includegraphics{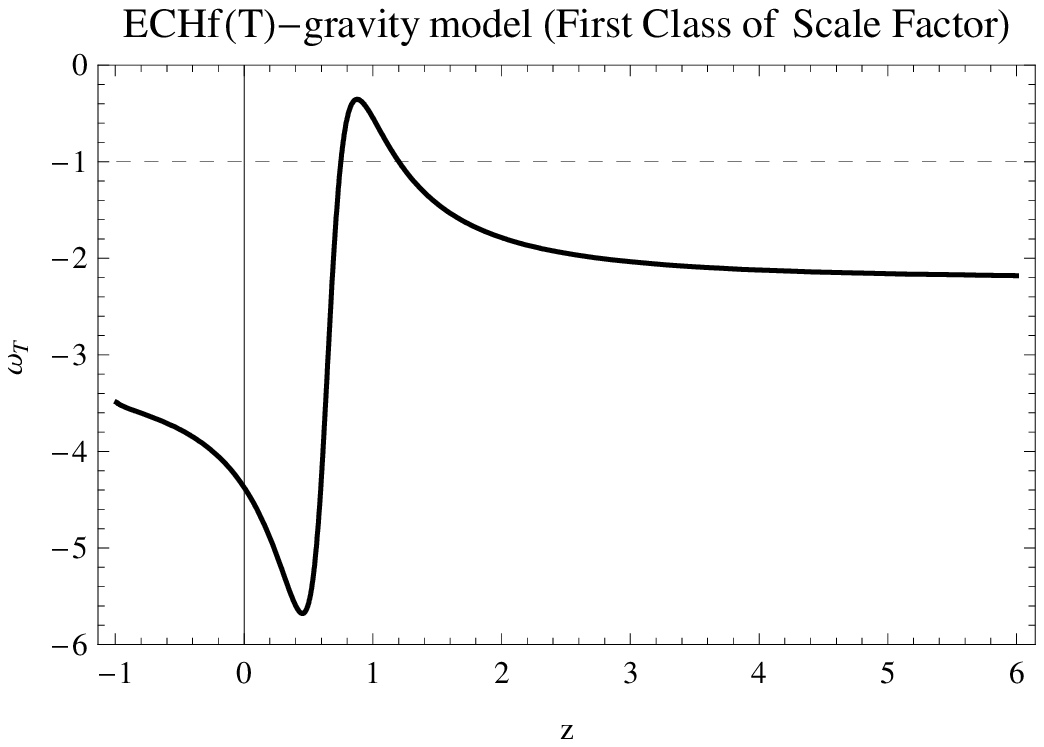}
      \vspace{5cm}
\caption[]{The EoS parameter of the entropy-corrected holographic
$f(T)$-gravity model, Eq. (\ref{wECHDE}), versus redshift for the
first class of scale factors, Eq. (\ref{a}). Auxiliary parameters
are: $c=0.818$ \cite{Li6}, $\alpha=-5$, $\beta=0.1$ and $h=0.55$.}
         \label{wT-ECHDE-Ph}
   \end{figure}

Note that the time-dependent EoS parameter (\ref{wECHDE}) in
contrast with constant EoS parameter (\ref{wHDE}) can justify the
transition from the quintessence state, $\omega_T>-1$, to the
phantom regime, $\omega_T<-1$, as indicated by recent observations
\cite{Sahni}. To illustrate this transition in ample detail, the EoS
parameter of the entropy-corrected holographic $f(T)$-gravity model,
Eq. (\ref{wECHDE}), versus redshift $z=\frac{a_0}{a}-1$ for the
first class of scale factors, Eq. (\ref{a}), is plotted in Fig.
\ref{wT-ECHDE-Ph}. Note that the torsion scalar $T$ can be expressed
in terms of redshift $z$. For the first class of scale factors (28)
one can obtain
$$T=-\frac{6h^2}{(t_s-t)^2}=-\frac{6h^2}{(1+z)^{2/h}}.$$ Figure \ref{wT-ECHDE-Ph} presents that for a
set of free parameters $c=0.818$ \cite{Li6}, $\alpha=-5$,
$\beta=0.1$ and $h=0.55$, $\omega_T$ crosses the $-1$ line twice. At
the transition redshift $z_T\simeq 0.75$, we have a direct
transition from $\omega_T>-1$ (quintessence phase) to $\omega_T<-1$
(phantom phase). Whereas at $z_T\simeq 1.20$, the crossing direction
is opposite, i.e. $\omega_T<-1\rightarrow\omega_T>-1$. Crossing the
$-1$ line twice in the direct and opposite transitions is in
agreement with that obtained recently for some $f(T)$-gravity models
\cite{Wu}.

Considering Eqs. (\ref{wECHDE}) and (\ref{ro ECH}) it seems that at
$T=\frac{\gamma}{\sigma+\delta\ln(-\lambda/T)}$, a singularity in
$\omega_T$ and a change of sign in $\rho_{\Lambda}$ appear.
Regarding $\omega_T$, Fig. \ref{wT-ECHDE-Ph} shows that the EoS
parameter of the entropy-corrected holographic $f(T)$-gravity model,
Eq. (\ref{wECHDE}), does not show any singularity.

\begin{figure}
\includegraphics{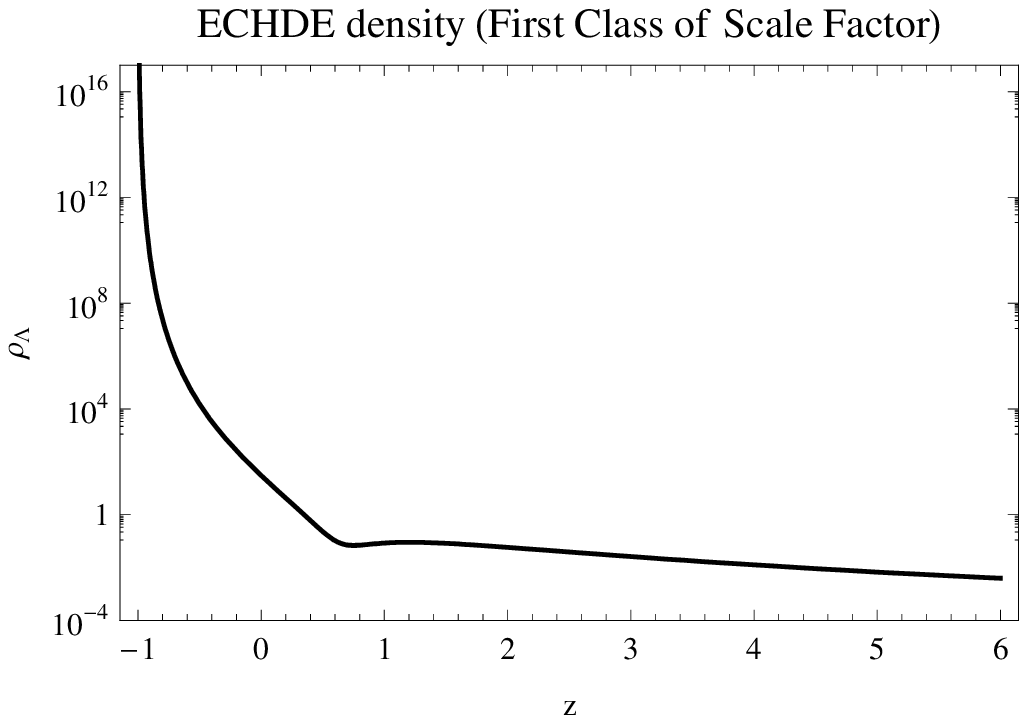}
      \vspace{5cm}
\caption[]{The ECHDE density, Eq. (\ref{ro ECH}), versus redshift
for the first class of scale factors, Eq. (\ref{a}). Auxiliary
parameters as in Fig. \ref{wT-ECHDE-Ph}.}
         \label{ECHDE-d-Ph}
   \end{figure}

To check the change of sign in $\rho_{\Lambda}$ given by Eq.
 (\ref{ro ECH}), we plot it in Fig. \ref{ECHDE-d-Ph}.
Figure \ref{ECHDE-d-Ph} clears that for the first class of scale
factors although a future Big Rip singularity in the ECHDE density
($\rho_{\Lambda}\rightarrow \infty$) occurs at $z\rightarrow -1$ (or
$t\rightarrow t_s$), the sign of $\rho_{\Lambda}$ does not change.
Also the EoS parameter remains finite at the future Big Rip
singularity when $z\rightarrow -1$ (see again Fig.
\ref{wT-ECHDE-Ph}). It is also interesting to note that Fig.
\ref{ECHDE-d-Ph} presents that the local minimum and maximum points
of $\rho_{\Lambda}$ occur at the transition redshifts when
$\omega_T=-1$ (see Fig. \ref{wT-ECHDE-Ph}). This can also be shown
analytically. From Eq. (\ref{roT}), ${\rm d}\rho_T/{\rm d}T=0$
yields
$$2Tf_{TT}+f_{T}-1=0.$$ Inserting the above relation into Eq. (\ref{omegaT}) gives $\omega_T=-1$.

\begin{figure}
\includegraphics{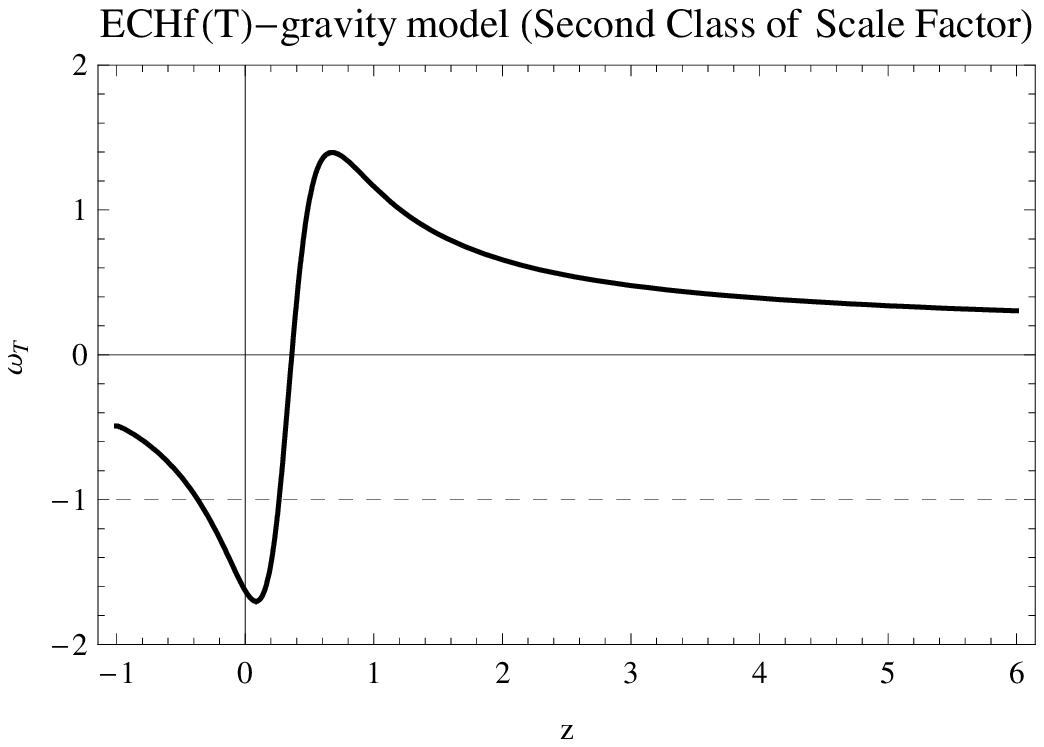}
      \vspace{5cm}
\caption[]{The EoS parameter of the entropy-corrected holographic
$f(T)$-gravity model, Eq. (\ref{wECHDEQ}), versus redshift for the
second class of scale factors, Eq. (\ref{aQ}). Auxiliary parameters
are: $c=0.818$ \cite{Li6}, $\alpha=-13$, $\beta=12$ and $h=1.31$.}
         \label{wT-ECHDE-Q}
   \end{figure}

\begin{figure}
\includegraphics{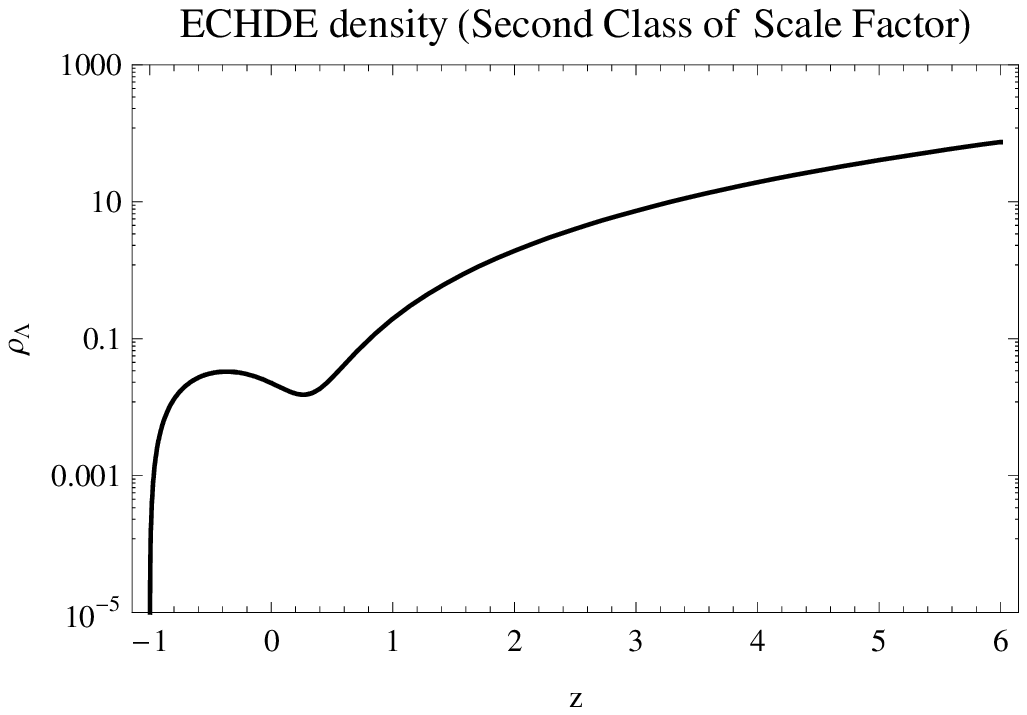}
      \vspace{5cm}
\caption[]{The ECHDE density, Eq. (\ref{ECHDE}), versus redshift for
the second class of scale factors, Eq. (\ref{aQ}). Auxiliary
parameters as in Fig. \ref{wT-ECHDE-Q}.}
         \label{ECHDE-d-Q}
   \end{figure}

For the second class of scale factors (\ref{aQ}), the resulting
$f(T)$ is the same as Eq. (\ref{fECHDE}) where
\begin{eqnarray}
\gamma=c^2{\left(\frac{h-1}{h}\right)}^2,
~~\delta=\frac{k^2\alpha}{18}{\left(\frac{h-1}{h}\right)}^4,
\nonumber\\
~~\lambda=\frac{6}{k^2}{\left(\frac{h}{h-1}\right)}^2,
~~\sigma=\frac{k^2\beta}{18}{\left(\frac{h-1}{h}\right)}^4.
\end{eqnarray}
Also the EoS parameter is obtained as
\begin{equation}
\omega_T=-1+\frac{2}{3h}\left[1+
\left({\frac{\delta-[\sigma+\delta\ln{(-\frac{\lambda}{T})}]}
{\gamma-[\sigma+\delta\ln{(-\frac{\lambda}{T})}]T}}\right)T\right],~~~h>1.\label{wECHDEQ}
\end{equation}
Here also in order to make $R_h$ be finite, the parameter $h$ should
be in the range of $h>1$. One notes that the dynamical EoS parameter
(\ref{wECHDEQ}) in contrast with the constant EoS parameter
(\ref{wHDEQ}) can accommodate the transition from $\omega_T>-1$ to
$\omega_T<-1$ at recent stage. Figure \ref{wT-ECHDE-Q} displays the
evolution of the EoS parameter of the entropy-corrected holographic
$f(T)$-gravity model, Eq. (\ref{wECHDEQ}), versus redshif $z$ for
the second class of scale factors, Eq. (\ref{aQ}). In this case, the
torsion scalar $T$ can be expressed in terms of redshift $z$ as
$$T=-\frac{6h^2}{t^2}=-6h^2(1+z)^{2/h}.$$ Figure \ref{wT-ECHDE-Q} like Fig. \ref{wT-ECHDE-Ph} shows that
the $-1$ line is crossed twice for another values set of the free
parameters, $c=0.818$ \cite{Li6}, $\alpha=-13$, $\beta=12$ and
$h=1.31$. At $z_T\simeq 0.26$ we have a direct transition (i.e.
$\omega_T>-1\rightarrow\omega_T<-1$). Also an opposite transition
occurs in the future at $z_T\simeq -0.37$. Furthermore, Fig.
\ref{wT-ECHDE-Q} clears that there is no any singularity in the
dynamical EoS parameter (\ref{wECHDEQ}). Note that also the sign of
the ECHDE density (\ref{ECHDE}) for the second class of scale
factors (\ref{aQ}) remains unchanged (see Fig. \ref{ECHDE-d-Q}).

\section{New agegraphic $f(T)$-gravity model}

For the NADE density \cite{Wei1}
\begin{equation}
\rho_{\Lambda}=\frac{3{n}^2}{k^2\eta^2},\label{NADE}
\end{equation}
the conformal time $\eta$ of the FRW universe is defined as
\begin{equation}
\eta=\int\frac{{\rm d}t}{a}=\int\frac{{\rm d}a}{Ha^2}.\label{eta}
\end{equation}
For the first class of scale factors (\ref{a}), the conformal time
$\eta$ by the help of Eq. (\ref{respect to r}) yields
\begin{equation}
\eta=\int_t^{t_s}\frac{{\rm
d}t}{a}=\frac{(t_s-t)^{h+1}}{a_0(h+1)}=\frac{h^{h+1}}{a_0(h+1)}{\left(\frac{-6}{T}\right)}^{\frac{h+1}{2}}.\label{E}
\end{equation}
Substituting Eq. (\ref{E}) into (\ref{NADE}) gives
\begin{equation}
\rho_\Lambda=\frac{\gamma}{2k^2}T^{h+1},\label{ro NEDE R}
\end{equation}
where
\begin{equation}
\gamma=\frac{6n^2a_0^2{(h+1)}^2}{{(-6h^2)}^{h+1}}.\label{alpha NADE}
\end{equation}
Equating (\ref{roT}) with (\ref{ro NEDE R}) yields
\begin{equation}
2Tf_T-f-T-\gamma T^{h+1}=0.\label{diff2}
\end{equation}
Solving Eq. (\ref{diff2}) results in the new agegraphic
$f(T)$-gravity model as
\begin{equation}
f(T)=\epsilon T^{1/2}+T+\frac{\gamma}{1+2h}T^{h+1} ,\label{fNADE}
\end{equation}
where $\epsilon$ is an integration constant. Inserting Eq.
(\ref{fNADE}) into (\ref{omegaT}) gives
\begin{equation}
\omega_{T}=-1-\frac{2(h+1)}{3h},~~~h>0,\label{wNADE}
\end{equation}
which is always smaller than $-1$ like the EoS parameter of the
holographic $f(T)$-gravity model (\ref{wHDE}), and it behaves as a
phantom type DE.

For the second class of scale factors (\ref{aQ}) and using
(\ref{respect to rQ}), the conformal time $\eta$ is obtained as
\begin{equation}
\eta=\int_0^t\frac{{\rm
d}t}{a}=\frac{t^{1-h}}{a_0(1-h)}=\frac{h^{1-h}}{a_0(1-h)}{\left(\frac{-6}{T}\right)}^{\frac{1-h}{2}},~~~0<h<1,\label{EQ}
\end{equation}
where the condition $h<1$ is necessary due to having a finite
conformal time. The resulting $f(T)$ is
\begin{equation}
f(T)=\epsilon T^{1/2}+T+\frac{\gamma}{1-2h}T^{1-h} ,\label{fNADEQ}
\end{equation}
where
\begin{equation}
\gamma=\frac{6n^2a_0^2{(1-h)}^2}{{(-6h^2)}^{1-h}}.\label{alpha
NADEQ}
\end{equation}
Also the EoS parameter of the new agegraphic $f(T)$-gravity model is
obtained as
\begin{equation}
\omega_{T}=-1+\frac{2(1-h)}{3h},~~~0<h<1,\label{wNADEQ}
\end{equation}
which shows a quintessence-like EoS parameter $\omega_T>-1$. Here in
order to have $-1<\omega_T<-1/3$, the parameter $h$ should be in the
range of $1/2<h<1$.
\section{Entropy-corrected new agegraphic $f(T)$-gravity model}

Here, we reconstruct the $f(T)$-gravity model corresponding to the
ECNADE density \cite{HW}
\begin{eqnarray}
\rho_{\Lambda} = \frac{3n^2}{k^2\eta^2} +
\frac{\alpha}{{\eta}^4}\ln{\left(\frac{{\eta}^2}{k^2}\right)} +
\frac{\beta}{\eta^4},\label{ECNADE}
\end{eqnarray}
which closely mimics to that of the ECHDE density (\ref{ECHDE}) and
$R_h$ is replaced with the conformal time $\eta$.

For the first class of scale factors (\ref{a}), substituting Eq.
(\ref{E}) into (\ref{ECNADE}) one can get
\begin{eqnarray}
\rho_\Lambda=\frac{\gamma}{2k^2}T^{h+1}+ \frac{1}{2k^2}
\left[\sigma+\delta\ln\left(\frac{\lambda}{T^{h+1}}\right)
\right]T^{2(h+1)},\label{ro ECNA}
\end{eqnarray}
where
\begin{eqnarray}
\gamma=\frac{6n^2{a_0}^2{(h+1)}^2}{{(-6h^2)}^{h+1}},
~~\delta=\frac{2k^2\alpha{a_0}^4{(h+1)}^4}{{(-6h^2)}^{2(h+1)}},
\nonumber\\~~\lambda=\frac{{(-6h^2)}^{h+1}}{k^2{a_0}^2{(h+1)}^2},
~~\sigma=\frac{2k^2\beta{a_0}^4{(h+1)}^4}{{(-6h^2)}^{2(h+1)}}.
\end{eqnarray}
Equating (\ref{roT}) with (\ref{ro ECNA}) gives
\begin{equation}
2Tf_T-f-T-\gamma T^{h+1}-\left[\sigma+\delta\ln\left(\frac{\lambda}
{T^{h+1}}\right)\right] T^{2(h+1)}=0.\label{dif eq4}
\end{equation}
Solving the differential equation (\ref{dif eq4}) one can obtain the
entropy-corrected new agegraphic $f(T)$-gravity model as
\begin{eqnarray}
f(T)=\epsilon T^{1/2}+T+\frac{\gamma}{1+2h}T^{h+1}
~~~~~~~~~~~~~~~~~~~~~~~~~~~~~~~~~\nonumber\\+
\frac{1}{3+4h}\left\{\sigma+\delta
\left[\frac{2(1+h)}{3+4h}+\ln\left(\frac{\lambda}
{T^{h+1}}\right)\right]\right\}T^{2(h+1)},\label{fECNADE}
\end{eqnarray}
where $\epsilon$ is an integration constant.

\begin{figure}
\includegraphics{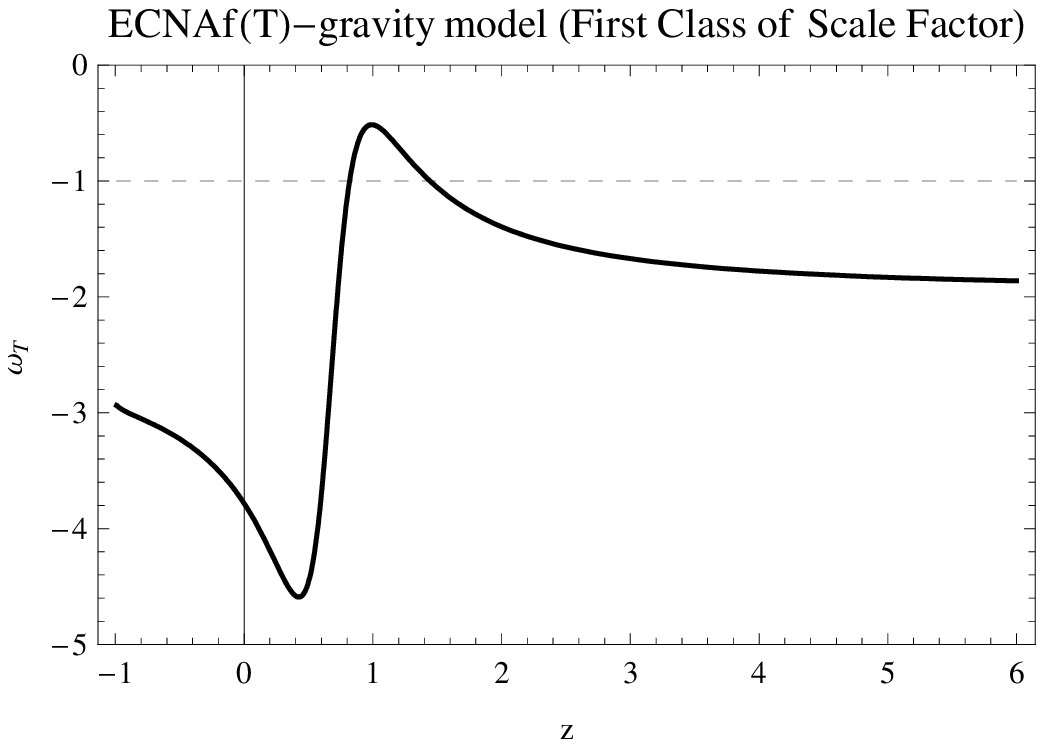}
      \vspace{5cm}
\caption[]{The EoS parameter of the entropy-corrected new agegraphic
$f(T)$-gravity model, Eq. (\ref{wECNADE}), versus redshift for the
first class of scale factors, Eq. (\ref{a}). Auxiliary parameters
are: $n=2.716$ \cite{Wei3}, $\alpha=-7.5$, $\beta=-14.8$ and
$h=2.5$.}
         \label{wT-ECNADE-Ph}
   \end{figure}

\begin{figure}
\includegraphics{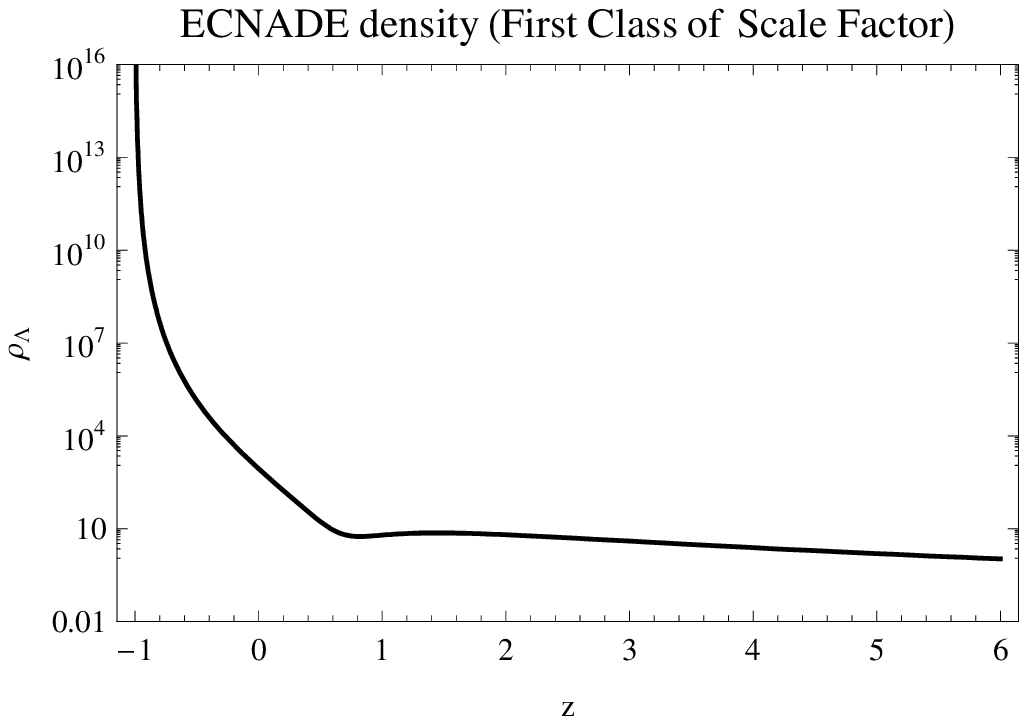}
      \vspace{5cm}
\caption[]{The ECNADE density, Eq. (\ref{ro ECNA}), versus redshift
for the first class of scale factors, Eq. (\ref{a}). Auxiliary
parameters as in Fig. \ref{wT-ECNADE-Ph}.}
         \label{ECNADE-d-Ph}
   \end{figure}

Inserting Eq. (\ref{fECNADE}) into (\ref{omegaT}) yields
\begin{equation}
\omega_T=-1-\frac{2}{3}\left(\frac{h+1}{h}\right)\left[1+
\left({\frac{-\delta+[\sigma+\delta\ln{(\frac{\lambda}{T^{h+1}})}]}
{\gamma+[\sigma+\delta\ln{(\frac{\lambda}{T^{h+1}})}]T^{h+1}}}\right)T^{h+1}\right],~~~h>0.\label{wECNADE}
\end{equation}
If we set $\delta=0=\alpha$ and $\sigma=0=\beta$ then Eqs.
(\ref{fECNADE}) and (\ref{wECNADE}) reduce to (\ref{fNADE}) and
(\ref{wNADE}), respectively. Note that the time-dependent EoS
parameter (\ref{wECNADE}) in contrast with constant EoS parameter
(\ref{wNADE}) can justify the transition from $\omega_T>-1$ to
$\omega_T<-1$. Figure \ref{wT-ECNADE-Ph} illustrates the EoS
parameter of the entropy-corrected new agegraphic $f(T)$-gravity
model, Eq. (\ref{wECNADE}), for the first class of scale factors,
Eq. (\ref{a}). Here for a values set of free parameters $n=2.716$
\cite{Wei3}, $\alpha=-7.5$, $\beta=-14.8$ and $h=2.5$, the direct
and opposite transitions occur at $z_T\simeq 0.82$ and $1.44$,
respectively. Besides, Fig. \ref{wT-ECNADE-Ph} reveals that there is
no any singularity in the dynamical EoS parameter (\ref{wECNADE}).
Note that here also the sign of the ECNADE density (\ref{ro ECNA})
for the first class of scale factors (\ref{a}) does not change (see
Fig. \ref{ECNADE-d-Ph}).

\begin{figure}
\includegraphics{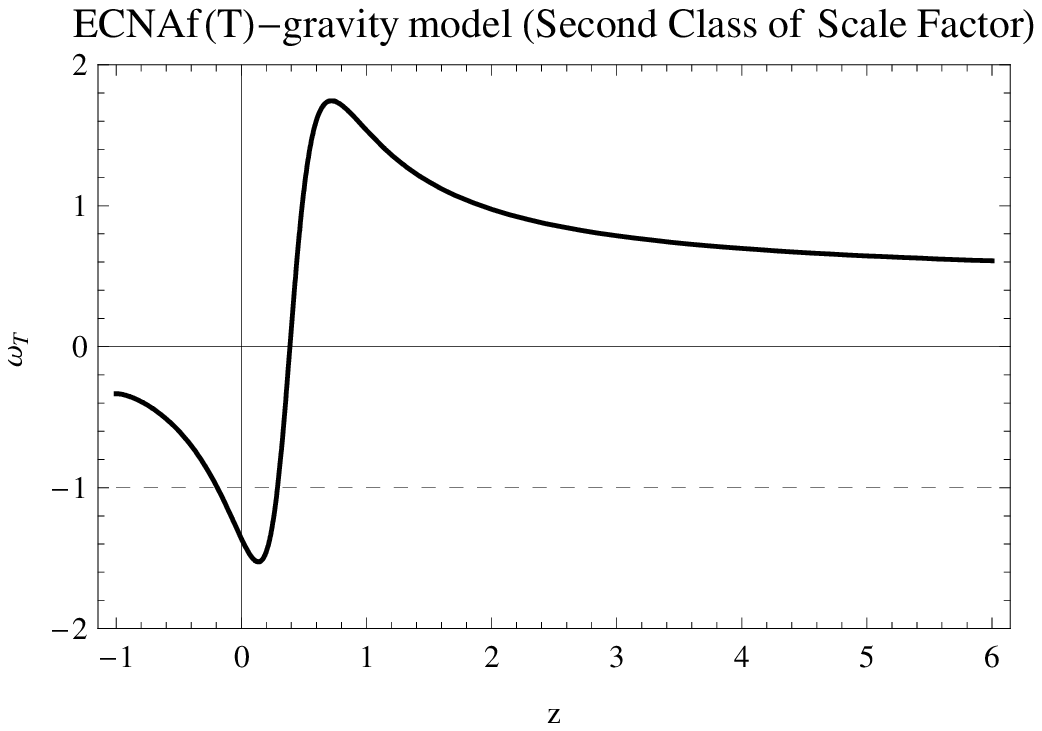}
      \vspace{5cm}
\caption[]{The EoS parameter of the entropy-corrected new agegraphic
$f(T)$-gravity model, Eq. (\ref{wECNADEQ}), versus redshift for the
second class of scale factors, Eq. (\ref{aQ}). Auxiliary parameters
are: $n=2.716$ \cite{Wei3}, $\alpha=-44$, $\beta=-10$ and $h=0.5$.}
         \label{wT-ECNADE-Q}
   \end{figure}

\begin{figure}
\includegraphics{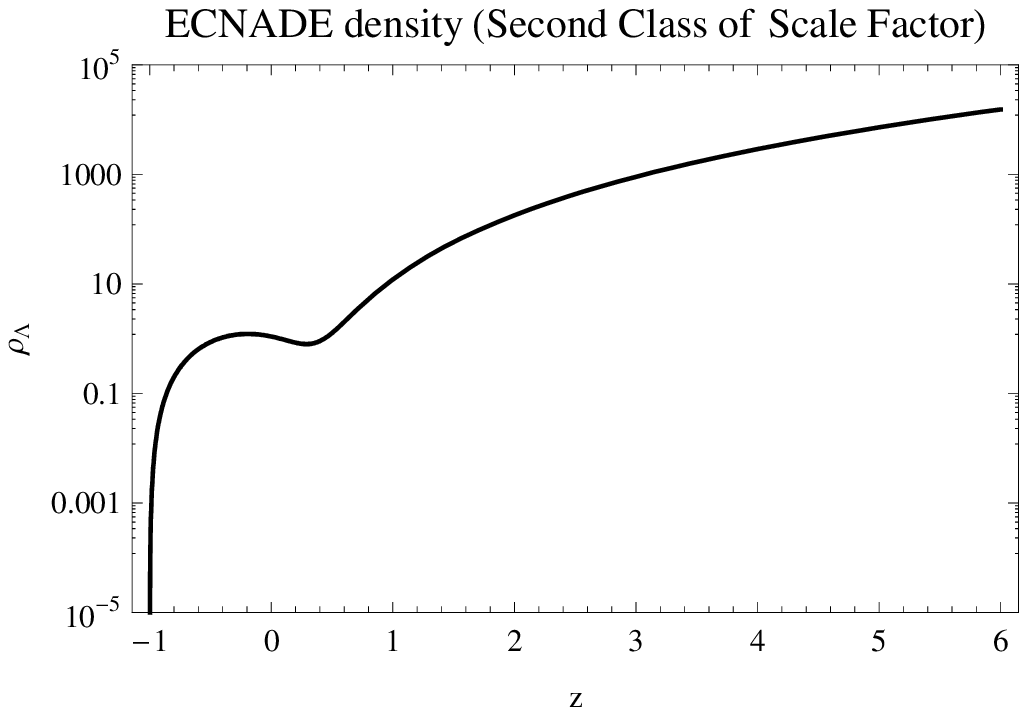}
      \vspace{5cm}
\caption[]{The ECNADE density, Eq. (\ref{ECNADE}), versus redshift
for the second class of scale factors, Eq. (\ref{aQ}). Auxiliary
parameters as in Fig. \ref{wT-ECNADE-Q}.}
         \label{ECNADE-d-Q}
   \end{figure}

For the second class of scale factors (\ref{aQ}), the resulting
$f(T)$ is the same as Eq. (\ref{fECNADE}) where
\begin{eqnarray}
\gamma=\frac{6n^2{a_0}^2{(1-h)}^2}{{(-6h^2)}^{1-h}},
~~\delta=\frac{2k^2\alpha{a_0}^4{(1-h)}^4}{{(-6h^2)}^{2(1-h)}},
\nonumber\\~~\lambda=\frac{{(-6h^2)}^{1-h}}{k^2{a_0}^2{(1-h)}^2},
~~\sigma=\frac{2k^2\beta{a_0}^4{(1-h)}^4}{{(-6h^2)}^{2(1-h)}}.
\end{eqnarray}
Also the EoS parameter can be obtained as
\begin{equation}
\omega_T=-1+\frac{2}{3}\left(\frac{1-h}{h}\right)\left[1+
\left({\frac{-\delta+[\sigma+\delta\ln{(\frac{\lambda}{T^{1-h}})}]}
{\gamma+[\sigma+\delta\ln{(\frac{\lambda}{T^{1-h}})}]T^{1-h}}}\right)T^{1-h}\right],~~~0<h<1.\label{wECNADEQ}
\end{equation}
Here also in order to have a finite conformal time $\eta$, the
parameter $h$ should be in the range of $0<h<1$. Contrary to the
constant EoS parameter (\ref{wNADEQ}), the dynamical EoS parameter
(\ref{wECNADEQ}) can accommodate the transition from $\omega_T>-1$
to $\omega_T<-1$ at recent stage. Figure \ref{wT-ECNADE-Q} presents
the evolution of the EoS parameter of the entropy-corrected new
agegraphic $f(T)$-gravity model, Eq. (\ref{wECNADEQ}), for the
second class of scale factors, Eq. (\ref{aQ}). Here also like Fig.
\ref{wT-ECNADE-Ph}, for another values set of the free parameters,
$n=2.716$ \cite{Wei3}, $\alpha=-44$, $\beta=-10$ and $h=0.5$,
$\omega_T$ crosses the $-1$ line twice at $z_T\simeq 0.29$ and
$-0.19$ corresponding to the direct and opposite transitions,
respectively. Besides, Fig. \ref{wT-ECNADE-Q} presents that there is
no any singularity in the dynamical EoS parameter (\ref{wECNADEQ}).
Note that here also the sign of the ECNADE density (\ref{ECNADE})
for the second class of scale factors (\ref{aQ}) does not change
(see Fig. \ref{ECNADE-d-Q}).

\section{Conclusions}
Here, we considered the original and entropy-corrected version of
the HDE and NADE models. Among various candidates explaining cosmic
accelerated expansion, only the HDE and NADE models are based on the
entropy-area relation. However, this definition can be modified by
the inclusion of quantum effects, motivated from the LQG. Hence the
ECHDE and ECNADE were introduced by addition of correction terms to
the energy densities of the HDE and NADE, respectively \cite{HW}.

We investigated the HDE, ECHDE, NADE and ECNADE in the framework of
$f(T)$-gravity. Among other approaches related with a variety of
 DE models, a very promising approach to DE is related
with the modified teleparallel gravity known as $f(T)$-gravity, in
which DE emerges from the modification of torsion. The class of
$f(T)$-gravity theories is an intriguing generalization of
Einstein's new GR, taking a curvature-free approach and using a
connection with torsion. It is analogous to the $f(R)$ extension of
the Einstein-Hilbert action of standard GR, but has the advantage of
the second order field equations \cite{linder}. We reconstructed
different theories of modified gravity based on the $f(T)$ action in
the spatially-flat FRW universe for two classes of scale factors
containing i) $a=a_0(t_s-t)^{-h}$ and ii) $a=a_0t^h$ and according
to the original and entropy-corrected version of the HDE and NADE
scenarios. Furthermore, we obtained the EoS parameters of the
corresponding $f(T)$-gravity models. Our calculations show that for
the first class of scale factors, the EoS parameters of the
holographic and new agegraphic $f(T)$-gravity models always behave
as that of phantom DE. Whereas for the second class, the EoS
parameters of the above-mentioned models behave like quintessence
EoS parameter. The EoS parameters of the entropy-corrected
holographic and new agegraphic $f(T)$-gravity models can cross the
phantom-divide line twice. For the first class of scale factors
$a=a_0(t_s-t)^{-h}$, the EoS parameters of both entropy-corrected
holographic and new agegraphic $f(T)$-gravity models have an
opposite transition ($\omega_T<-1\rightarrow\omega_T>-1$) in the far
past and a direct transition ( $\omega_T>-1\rightarrow\omega_T<-1$)
in the near past. For the second class of scale factors $a=a_0t^h$,
the EoS parameters of both entropy-corrected holographic and new
agegraphic $f(T)$-gravity models have a direct transition in the
near past and an opposite transition in the future. It is
interesting to note that the direct transition from the non-phantom
phase to the phantom one in the near past is consistent with the
recent cosmological observational data \cite{Sahni}.
\\
\\
\noindent{{\bf Acknowledgements}}\\
The authors thank Dr. Arash Sorouri and Dr. Mubasher Jamil for
helping in good English presentation of the paper. The work of K.
Karami has been supported financially by Research Institute for
Astronomy $\&$ Astrophysics of Maragha (RIAAM), Maragha, Iran.

\clearpage

\clearpage

\clearpage

\clearpage
\end{document}